\documentclass[aps,pra,twocolumn,groupedaddress,nofootinbib,longbibliography]{revtex4-1}
\usepackage{amsmath}
\usepackage{amssymb}
\usepackage{graphicx}
\usepackage{mathrsfs}
\usepackage{ntheorem}
\usepackage{mathtools}
\usepackage{enumerate}
\usepackage{enumitem}
\usepackage[T1]{fontenc}
\usepackage{physics}
\usepackage{subfigure}
\usepackage{overpic}
\usepackage{epstopdf}
\usepackage{bm}
\usepackage{color,soul}
\usepackage[bookmarks=false]{hyperref}
\usepackage{xcolor}
\hypersetup{colorlinks=true,citecolor=blue,
linkcolor=blue,urlcolor=blue,pdfstartview=FitH,
bookmarksopen=true}
\newcommand{\ot}{\otimes}	
\newcommand{\kethjh}[1]{|{#1}\rangle}
\newcommand{\brahjh}[1]{\langle{#1}|}
\newcommand{\bkt}[2]{\langle{#1}|{#2}\rangle}
\begin{document}
\title{Improving the precision of weak-value-amplification with two cascaded Michelson  interferometers based on Vernier-effect}

\author{Jing-Hui Huang$^{1}$}\email{Email:jinghuihuang@cug.edu.cn}
\author{Fei-Fan He$^{1}$}
\email{Email:hfeifan2017@cug.edu.cn}
\author{Xue-Ying Duan$^{2,3,4}$} \email{Email:CUGxyDuan@163.com}
\author{Guang-Jun Wang$^{2,3,4}$} \email{Email:gjwang@cug.edu.cn}
\author{Xiang-Yun Hu$^{1}$} \email{Email:xyhu@cug.edu.cn}

\address{$^{1}$ School of Institute of Geophysics and Geomatics, China University of Geosciences, Lumo Road 388, 430074 Wuhan, China. }
\address{$^{2}$ School of Automation, China University of Geosciences, Lumo Road 388, 430074 Wuhan, China.}
\address{$^{3}$ Hubei Key Laboratory of Advanced Control and Intelligent Automation for Complex Systems, China }
\address{$^{4}$ Engineering Research Center of Intelligent Technology for Geo-Exploration, Ministry of Education, China}

\begin{abstract}
A modified-weak-value-amplification(MWVA) technique of measuring the mirror's velocity  based on the Vernier-effect has been proposed. We have demonstrated with  sensitivity-enhanced and the higher signal-to-noise ratio(${\rm SNR}$) by using two cascaded Michelson interferometers. These two interferometers are composed of similar optical structures. One interferometer with a fixed mirror acts as a fixed part of the Vernier-scale, while the other with a moving mirror acts as a sliding part of the Vernier-scale for velocity sensing. The envelope of the cascaded interferometers shifts much more than a single one with a certain enhancement factor, which is related to the free space range difference between these two interferometers. In addition, we calculate the ${\rm SNR}$ based on the Fisher information with both the MWVA technique and the traditional-weak-value-amplification(TMVA) technique. The results show that the ${\rm SNR}$ with our MWVA technique is larger than the the ${\rm SNR}$ with the TWVA technique within the range of our time measurement window. Our numerical analysis proved that our MWVA technique is more efficient than the TWVA technique. And by using the principles of the Vernier-effect, it is applicative and convenient to ulteriorly improving the sensitivity and ${\rm SNR}$ in measuring other quantities with the MWVA technique.
\end{abstract}
\maketitle

\section{Introduction}
\label{intro}
A weak measurement is a new tool for characterizing post-selected quantum systems in the recent development of quantum mechanics\cite{PhysRevLett.126.020502,PhysRevLett.126.100403}. 
Historically speaking, the idea of this measurement was originally proposed by Aharonov, Albert and Vaidman\cite{AAV} as an extension to the standard("projective") von-Neumann model of measurement. In weak measurement, information is gained by weakly probing the system, while approximately preserving its initial state. The average shift of the final pointer state can go far beyond the eigenvalue spectrum of the system observable in sharp contrast to the standard quantum measurement. This average shift is called the weak value, which is usually complex\cite{PhysRevLett.62.2326,PhysRevLett.62.2325}. These features have allowed a wide range of applicability in understanding many counter-intuitive quantum results. For example, weak value can offer new insights into quantum paradox, such as Hardy's paradox\cite{PhysRevLett.102.020404,Yokota_2009}, the three-box paradox\cite{RESCH2004125}, apparent superluminal travel\cite{PhysRevA.66.042102}, direct measuring the real and imaginary components of the wave-function \cite{2011Spectroscopy}. 

In addition to the fundamental physics interest in weak values, one of the major goals in weak measurement is to enhance the sensitivity of estimating weak signals\cite{2011Mechanical,PhysRevLett.112.200401,2014Amplification,2020Anomalous,Huang2021}. For example, Dixon et al. amplified very small transverse deflections of an optical beam, then measured the angular deflection of a mirror down to 400 $\pm$ 200 frad and the linear travel of a piezo actuator down to 14 $\pm$7 fm \cite{2009Ultrasensitive}. Boyd et al. demonstrated the first realization of weak-value amplification in the azimuthal degree of freedom and had achieved effective amplification factors as large as 100\cite{PhysRevLett.112.200401}. Viza et al. achieved a velocity measurement of 400 fm/s by measuring one Doppler shifted due to a moving mirror in a Michelson interferometer\cite{2013Weak}. Pati et al. even proposed an alternative method to measure the temperature of a bath using the weak measurement scheme with a finite-dimensional meter\cite{PhysRevA.102.012204}. Therefore, these applications have been termed as the  weak-value-amplification(WVA) technique in literature\cite{PhysRevA.102.042601}.

It is worth noting that weak-value-amplification cannot be arbitrarily large in the cost of decreasing output laser intensity\cite{PhysRevLett.115.120401}. The analysis of Koike et al.\cite{PhysRevA.84.062106} showed that the measured displacement and the amplification factor are  not proportional to the weak value and rather vanish in the limit of infinitesimal output intensity. The work of Pang et al.\cite{PhysRevA.90.012108} investigated the limiting case of continuous-variable systems to demonstrate the influence of system dimension on the amplification limit. Therefore, it is still the focus in weak measurement to enhance the sensitivity of detecting small signals. The experiment of Starling et al.\cite{PhysRevA.80.041803}  and the proposal of Feizpour et al. \cite{PhysRevLett.107.133603} showed that post-selection can significantly raise the signal-to-noise ratio (${\rm SNR}$) of weak measurement. Huang et al\cite{PhysRevA.100.012109} proposed an approach called dual weak-value amplification (DWVA), with sensitivity several orders of magnitude higher than the standard approach without losing signal intensity.  Krafczyk et al\cite{PhysRevLett.126.220801}. experimentally demonstrated recycled weak-value measurements. By using photon counting detectors, they demonstrated a signal improvement by a factor of $4.4 \pm 0.2$ and a signal-to-noise ratio improvement of $2.1 \pm 0.06$, compared to a single-pass weak-value experiment.

Recently, Vernier-effect has been an effective tool to enhance the sensitivity of photonic devices\cite{Dai:09,Claes10,Zhu:17,Liu:17}. The Vernier-effect is an efficient method to enhance the accuracy of measurement instruments, which consists of two/three scales with different periods.  In Ref.\cite{Dai:09} a sensor is introduced that consists of two cascaded ring resonators, and it is shown theoretically that it can obtain very high sensitivities thanks to the Vernier principle. In \cite{Zhu:17} a high sensitivity of 1892dB/RIU of the optical sensor was achieved for the intensity interrogation based on the cascaded reflective Mach–Zehnder interferometer(MZI)s and the micro-ring resonators. Further, in Ref.\cite{Liu:17} a three cascaded micro-ring resonators based refractive index optical sensor with a high sensitivity of 5866 nm/RIU was demonstrated, the measurement range of which is significantly improved 24.7 times compared with the traditional two cascaded micro-ring resonators.

In this letter, we put forward a modified-weak-value-amplification(MWVA) with two cascaded Michelson interferometers to improve the sensitivity of traditional-weak-value-amplification(TMVA) based on the Vernier-effect. More specifically, in the framework of quantum weak measurements, we observe the temporal shift induced by a small spectral shift with two cascaded Michelson interferometers. The spectral shift in our scheme is a Doppler frequency shift produced by a mirror in one of the interferometers, while the other interferometer has the same optical structure with no movement of the mirror. By numerical simulation, We obtain the temporal shift dependence of the velocity, the sensitivity, and the signal-to-noise ratio (${\rm SNR}$) calculated from the Fisher information\cite{2010Quantum}. Besides, our MWVA technique is comparable to the TWVA technique, but allows us to reach the higher sensitivity and ${\rm SNR}$ under the same conditions of measuring time.

The paper is organized as follows. In Sec II.A, we briefly review the TWVA technique for measurement velocity in a single Michelson interferometer. Then, in Sec II.B, we first derive the MWVA technique for the velocity with two cascaded Michelson interferometers based on Vernier-effect. In Sec III, by ensuring the same time measurement window and determine the respective meter in the two techniques, we obtained analytic results of the sensitivity and the ${\rm SNR}$ from the Fisher information. Sec IV is devoted to the summary and the discussions.

\section{The TWVA technique and the MWVA technique} 
In this article, we propose and numerically demonstrate a weak measurement scheme, in which the amplification of the phase shifts in a Michelson interferometer can be effectively enhanced by introducing another cascaded interferometer based on the Vernier-effect. In order to compare our new technique to the traditional ones, we briefly review the traditional WVA technique to measuring velocity with a Michelson interferometer by following the Ref.\cite{2013Weak} in the following subsection.

\subsection{The WVA technique for velocity measurements} 

\begin{figure}[t]
\vspace*{-5mm}
\centering \includegraphics[width=0.45\textwidth]{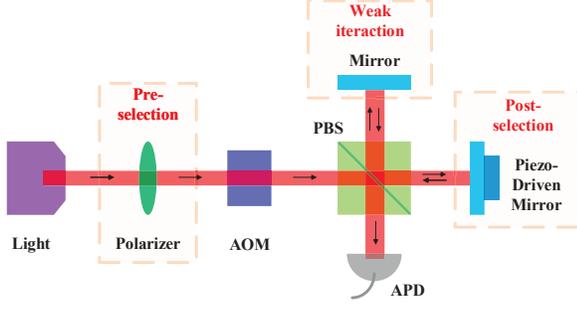}
\caption{Schematic of the WVA technique for velocity measurements in the single Michelson. AOM is the Acousto-Optic Modulator, which can modulate the light into a temporal Gaussian shaped pulse. PBS is the polarizer beam splitter. The arrival time of single photons is measured with an avalanche photodiode(APD).}
\label{Fig:singel-interferometer}
\end{figure}

In this section, we briefly review an interferometer scheme in Ref.\cite{2013Weak} combined WVA technique to measure longitudinal velocities with the temporal meter. The temporal shift is proportional to the weak value and can be amplified in the measurement of $v$, which is accompanied by a decrease in the measured intensity due to the nature of the weak measurement. The temporal shift is induced by the spectral shift, which is a Doppler frequency shift produced by a moving mirror in a Michelson interferometer. The weak measurement is characterized by state preparation, a weak perturbation, and post-selection. The detail of the WVA technique for velocity measurements is shown in Fig. \ref{Fig:singel-interferometer}. 

The TWVA technique of weak measurement can be characterized into three parts: state preparation, weak interaction, and post-selection. The initial state $\kethjh{\Phi_{i}}$ of the system and $\kethjh{\Psi_{i}}$ of the meter are prepared with the polarizer and AO Modulator. The initial polarization state of the system can be described by the polarization qubit:
\begin{eqnarray}
\label{inter_sy_initial}
\kethjh{\Phi_{i}}= sin( \alpha)\kethjh{H}+ cos(\alpha)\kethjh{V}
\end{eqnarray}
where $\alpha=\pi /4$ is the angle between the horizontal line and the transmission axis of line polarizer(pre-selection); The horizontal polarization state $\kethjh{H}$ correspond to the arm with the mirror which the light goes through, while the vertical polarization state $\kethjh{V}$ correspond to the arm with piezo-driven mirror in the Michelson interferometer. The preparation of the meter consists of the generation of the temporal mode 
\begin{eqnarray}
\label{inter_meter_initial}
I_{single}^{i}=|\bkt{p}{\Psi_{i}}|^{2} (t)= \frac{1}{\sqrt{2 \pi \tau ^{2}}}
e^{-(t+t_{0})^{2}/2\tau ^{2}}
\end{eqnarray}
where $\tau$ is the length of the Gaussian pulse and $t_{0}$ is the center of the pulse. The advantage of that temporal meter and the non-Fourier limited Gaussian-shaped pulse has been studied in Ref.\cite{2013Weak}. The pulse is injected into a
Michelson interferometer, where the horizontally polarized component of the pulse goes through the arm with a slowly moving mirror at speed $v$, and the vertically one goes through the arm with a piezo-Driven mirror. The weak interaction in Fig. \ref{Fig:singel-interferometer} can be expressed as 
\begin{eqnarray}
\label{inter_interaction}
U=e^{-ig\hat{A} \ot \hat{p}}=e^{-i \omega_{d} \hat{A}   t/2}
\end{eqnarray}
where $\omega_{d}=2 \pi f_{d}=2 \pi \frac{2v}{\lambda}$. Note that the spectral shift $f_{d}=2v/ \lambda$ is proportional to velocity $v$. The observable $\hat{A}$ satisfies:
$\hat{A}= \frac{1}{2}(\kethjh{H} \brahjh{H}-\kethjh{V} \brahjh{V})$.

\begin{figure}[b]
\centering 
\includegraphics[width=0.5\textwidth]{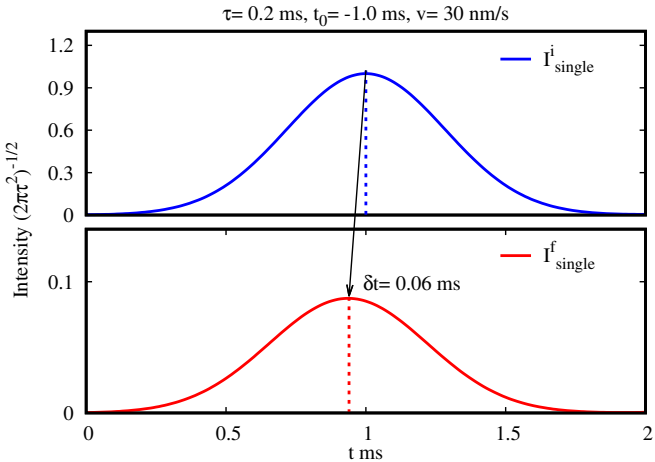}
\caption{The time shift of meter in the scheme of the WVA technique for measuring velocity $v=30 $ nm/s with the length of the Gaussian pulse $\tau=$ 0.2 ms and the phase offset $\phi$= 0.3 rad of the post-selection. The vertical dot line represents the corresponding the center of the Gaussian function.}
\label{Fig:result-singel-interferometer}
\end{figure}

Finally, the post-selection of the weak measurement is controlled by inducing a phase offset $2\phi$ by the piezo-Driven mirror. Thus, the final post-selection of the system is given by:
\begin{eqnarray}
\label{inter_sy_final}
\kethjh{\Phi_{f}}= \frac{1}{\sqrt{2}}(\kethjh{H} -e^{i2\phi}\kethjh{V})
\end{eqnarray}
and the weak value can be calculated by:
\begin{eqnarray}
\label{weak_value}
A_{w}=\frac{\brahjh{\Phi_{f}}\hat{A}\kethjh{\Phi_{i}}}{\bkt{\Phi_{f}}{\Phi_{i}}}=-i \rm cot \phi \approx \frac{-i}{\phi},
\end{eqnarray}
By following the original work in Ref.\cite{AAV}, the meter state of the temporal mode  after the post-selection becomes:
\begin{eqnarray}
\bkt{p}{\Psi_{f}} &=& \brahjh{\Phi_{f}} U \kethjh{\Phi_{i}} \bkt{p}{\Psi_{i}}
\nonumber \\
&=& \brahjh{\Phi_{f}} e^{-ig\hat{A} \ot \hat{p}}  \kethjh{\Phi_{i}} \bkt{p}{\Psi_{i}} 
\nonumber \\
&=& \bkt{\Phi_{f}}{\Phi_{i}} [1-igA_{w}\hat{p}]   \bkt{p}{\Psi_{i}} +O(g^{2})
\nonumber \\
&\approx&\bkt{\Phi_{f}}{\Phi_{i}}e^{-igA_{w}\hat{p}}\bkt{p}{\Psi_{i}} 
\label{Eq:the_meter_state}
\end{eqnarray}
and therefore, the absolute value squared of the meter state (\ref{Eq:the_meter_state}) is given by:
\begin{eqnarray}
\label{inter_meter_final2}
I_{single}^{f}&=& |\bkt{p}{\Psi_{f}}|^{2} \nonumber \\
&=&(sin\phi)^{2}e^{-\frac{2\pi v t}{\lambda \phi} } |\bkt{p}{\Psi_{i}}|^{2} \nonumber \\
&=& \frac{(sin\phi)^{2}}{\sqrt{2 \pi \tau ^{2}}}
e^{-\frac{2\pi v t}{\lambda \phi}} e^{-(t+t_{0})^{2}/2\tau ^{2}} \nonumber \\
&\approx& \frac{(sin\phi)^{2}}{\sqrt{2 \pi \tau ^{2}}}
e^{-(t+t_{0}+ \frac{4\pi \tau ^{2} v}{\lambda \phi})^{2}/2\tau ^{2}}
\end{eqnarray}
The final step approximation of Eq.(\ref{inter_meter_final2}) condition is $4\pi v \tau ^{2} \ll \lambda \phi$. In the scheme of the traditional WVA technique for measuring velocity, the time shift $\delta t =\frac{4\pi \tau ^{2} v}{\lambda \phi}$ in Eq.(\ref{inter_meter_final2}) is amplified in the measurement of $v$. Note that the work in Ref.\cite{2013Weak} and our derivation show that the spectral shift $f_{d}=2v/ \lambda$ can translate into the time shift $\delta t =\frac{2\pi \tau ^{2} f_{d}}{ \phi}=\frac{4\pi \tau ^{2} v}{\lambda \phi}$.

In addition, we show the time shift of meter in the scheme of the TWVA technique for measuring velocity $v=30 $ nm/s with the length $\tau=$ 0.2 ms and central wavelength $\lambda=$ 800 nm of the Gaussian pulse in Fig.  \ref{Fig:result-singel-interferometer}. Note that the work in Ref.\cite{PhysRevLett.105.010405} studied an interferometric scheme based on a purely imaginary weak value, combined with a frequency-domain analysis, which may have the potential to outperform standard interferometry by several orders of magnitude. However, the amplification of the TWVA technique can not be arbitrarily large due to the limit of the resolution of the detector. Therefore, the MWVA technique based on Vernier-effect  with two cascaded Michelson interferometers for measuring velocity is proposed in the next subsection.

\subsection{The MWVA technique based on  Vernier-effect} 

\begin{figure}[t]
\centering \includegraphics[width=0.49\textwidth]{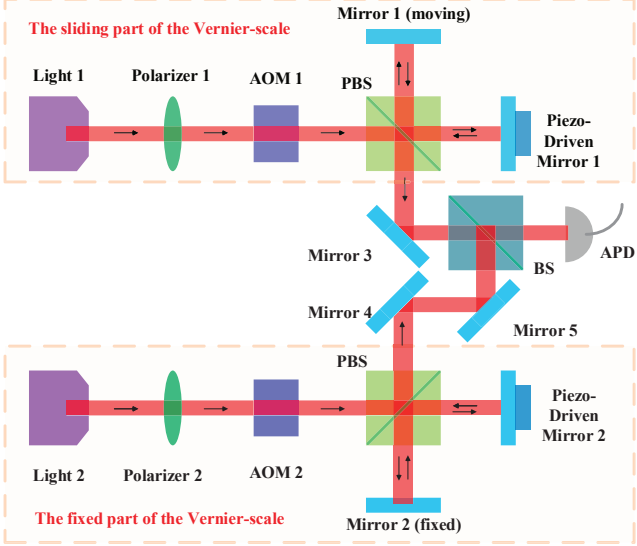}
\caption{The scheme of the MWVA technique with two cascaded Michelson interferometers based on Vernier-effect.}
\label{Fig:two-interferometer}
\end{figure}

In this section, we propose a sensitivity-enhanced scheme of the MWVA technique with two cascaded Michelson interferometers based on the Vernier-effect. In addition, through numerical simulation experiments, we verify the effectiveness and feasibility of the MWVA technique. 

The schematic diagram of the MWVA technique for measuring velocity is shown in Fig. \ref{Fig:two-interferometer}. It includes two Michelson interferometers based on the TWVA technique, and each Michelson interferometer is the same as the Michelson interferometer with the TWVA technique in Fig. \ref{Fig:singel-interferometer}. The upper interferometer with moving mirror 1 serves as the sliding part of the Vernier-scalar, as the velocity changes will cause a shift of the final meter. The lower interferometer with statical mirror 2 serves as the fixed part of the Vernier-scalar. The Vernier-scale is an efficient method to enhance the accuracy of measurement instruments, which consists of two scales with different periods\cite{SHAO201573}. 

In our work, the upper interferometer and the lower interferometer are prepared with different free spectrum range(FSR). The total transmission meter of the cascaded Michelson interferometer is the superposition of the optical power output of two interferometers, which exhibits peaks at times where two interference peaks of the respective interferometers partially overlap. The envelope period is given by
\begin{eqnarray}
\label{envelope_period}
\frac{FSR_{sliding} \times FSR_{fixed}}
{|FSR_{sliding} - FSR_{fixed}|}
\end{eqnarray}
When the velocity change,  the transmission of the single(upper) interferometer will shift. Then the shift of cascaded Michelson interferometers envelope is magnified by a certain factor $Enh$. The shift of the transmission peak is proportional to the enhanced factor
\begin{eqnarray}
\label{enhanced_factor}
 {Enh}=\frac{ FSR_{fixed}}
{|FSR_{sliding} - FSR_{fixed}|}
\end{eqnarray}
In order to produce the Vernier-effect, the parameters of the main optical components in our scheme in Fig. \ref{Fig:two-interferometer} shall meet the following requirements:

(i) \textbf{Light}: high-power light source is suitable for our scheme. Note that the pulses for light 1 and light 2 don't have to be coherent, because the laser detected at APD is the superposition of the optical power output of two interferometers.

(ii) \textbf{Piezo
-Driven Mirror}: the post-selection of the TWVA technique is controlled by inducing a phase offset $2 \phi$  with the Piezo-Driven Mirror. In other words, the intensity of the outgoing light is proportional to the phase offset $2 \phi$, which is shown in Eq.(\ref{inter_meter_final2}). To obtain the overlap with easily discernible peaks, the intensity of the outgoing light of each interferometer, namely the phase offset $2 \phi$ should keep the same value.

(iii) \textbf{AOM}: AOM is the most critical device in our scheme. According to the researches in Ref. \cite{SHAO201573,Zhu:17,Liu:17}, the pulse, which serves as the sliding part or the fixed part of the Vernier-scale, ought to be produced with 
equally spaced Gaussian mode rather than individual pulse (\ref{inter_meter_initial}). At the same time, AOM is mainly used outside the laser cavity, using the electric signal of the driving source to modulate the laser\cite{Zeng2007Acousto}. In our work, AOM 1  modulates the laser with $FRS_{sliding}$ as the pre-selection meter $|\bkt{p}{\Psi_{i}}|_{sliding}^{2} (t)$ $=I_{u}^{i}$ in upper interferometer, while AOM 2  modulates the laser with $FRS_{fixed}$ as the pre-selection meter $|\bkt{p}{\Psi_{i}}|_{fixed}^{2} (t)$ $=I_{d}^{i}$ in lower interferometer. Then $I_{u}^{i}$ and $I_{d}^{i}$ are given by:
\begin{eqnarray}
\label{MVP_meter_initial_sliding}
I_{u}^{i} 
=\sum_{m=0}^{N}
\frac{1}{\sqrt{2 \pi \tau ^{2}}}
e^{-(t+t_{0}-m \times FRS_{sliding})^{2}/2\tau ^{2}}
\end{eqnarray}
\begin{eqnarray}
\label{MVP_meter_initial_fixed}
I_{d}^{i} 
=\sum_{m=0}^{N}
\frac{1}{\sqrt{2 \pi \tau ^{2}}}
e^{-(t+t_{0}-m \times FRS_{fixed})^{2}/2\tau ^{2}}
\end{eqnarray}

The final meter $|\bkt{p}{\Psi_{f}}|_{fixed}^{2} (t)$ =$I_{d}^{f}$ of the lower interferometer becomes:
\begin{eqnarray}
\label{MVP_meter_final_fixed}
I_{d}^{f} 
=\sum_{m=0}^{N}
\frac{(sin\phi)^{2}}{\sqrt{2 \pi \tau ^{2}}}
e^{-(t+t_{0}-m \times FRS_{fixed})^{2}/2\tau ^{2}}
\end{eqnarray}

The final meter $|\bkt{p}{\Psi_{f}}|_{sliding}^{2} (t)$=$I_{u}^{f}$ of the upper interferometer with moving mirror 1 at velocity $v$ becomes:
\begin{eqnarray}
\label{MVP_meter_final_sliding}
I_{u}^{f} 
=\sum_{m=0}^{N}
\frac{(sin\phi)^{2}}{\sqrt{2 \pi \tau ^{2}}}
e^{-(t+t_{0}-m \times FRS_{sliding} + \frac{4\pi \tau ^{2} v}{\lambda \phi})^{2}/2\tau ^{2}}
\end{eqnarray}
The intensity $I_{sum}^{f}=I_{u}^{f}+I_{d}^{f}$ of the total laser  detected at APD is the sum of the light power $I_{u}^{f} $ from the upper interferometer and the light power $I_{d}^{f}$ from the lower interferometer.

\begin{figure*}[t]
	\centering
\subfigure
{
	\vspace{-0.2cm}
	\begin{minipage}{9.2cm}
	\centering
	\centerline{\includegraphics[scale=0.699,angle=0]{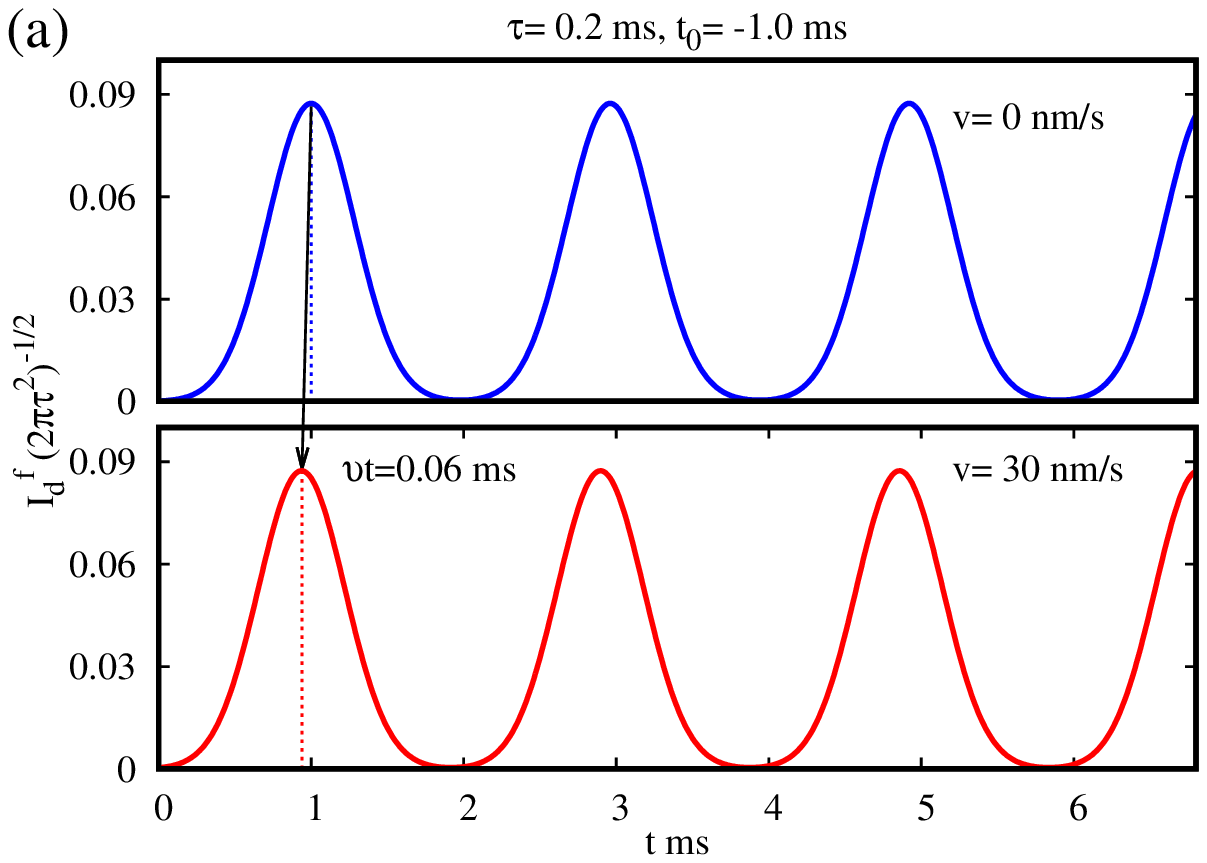}}
	\end{minipage}
}
\vspace{-0.2cm}
\subfigure
{
	\begin{minipage}{8cm}
	\centering
	\centerline{\includegraphics[scale=0.699,angle=0]{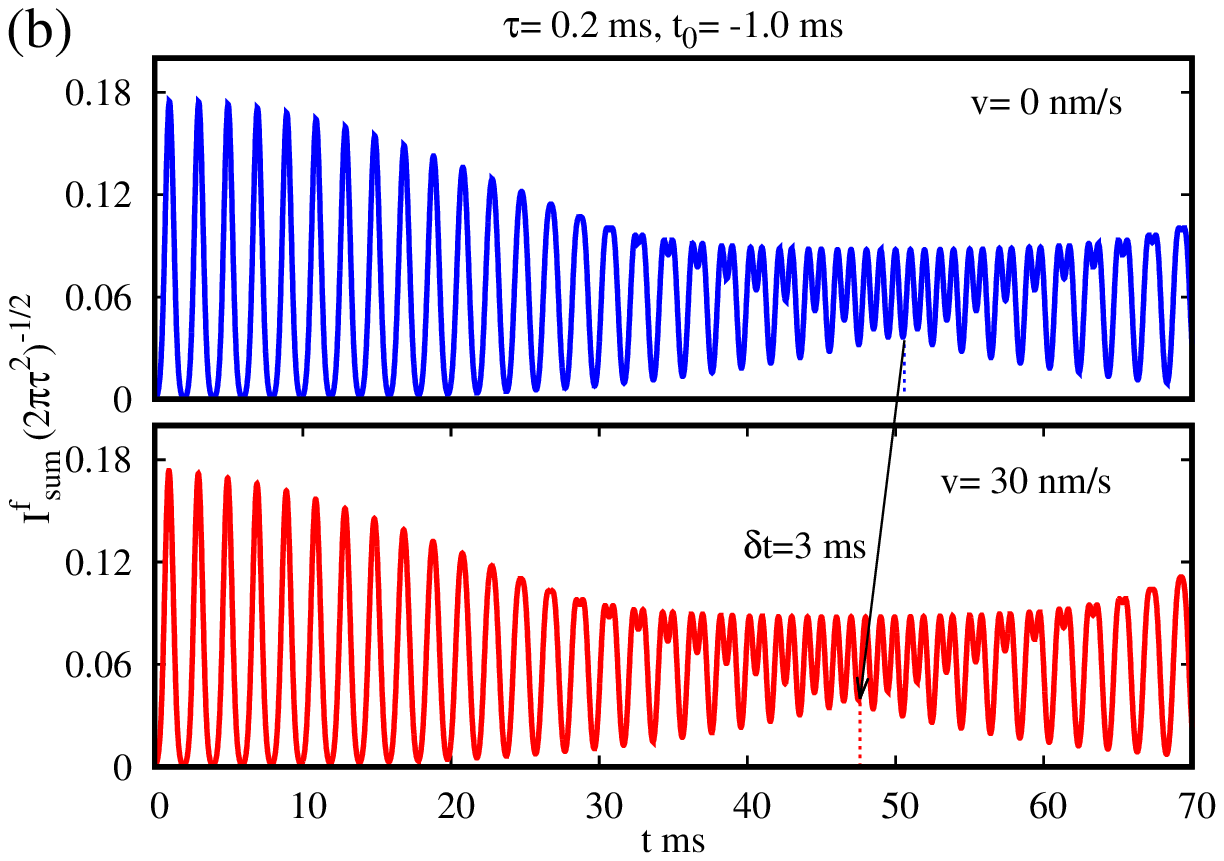}}
	\end{minipage}
}
\vspace{-0.2cm}
\vspace*{0mm} \caption{\label{Fig:result-many-two-interferometer} 
Numerical simulation of the transmission temporal shift of (a) single Michelson interferometer and (b) two cascaded Michelson interferometers.}
\end{figure*}

In the end, we fix a specific experimental setup, that is, the given chosen post-selection phase $\phi=0.3  $ rad, the length $\tau=$ 0.2 ms and central wavelength $\lambda=$ 800 nm, the parameter $t_{0}$=0.0 of the pulse $I_{u}^{i} $ (\ref{MVP_meter_initial_sliding}) and the pulse $I_{d}^{i} $ (\ref{MVP_meter_initial_fixed}), the free spectrum range $FSR_{fixed}$=2.0 ms of the lower interferometer and the free spectrum range $FSR_{sliding}$=1.96 ms of the upper interferometer in Fig. \ref{Fig:two-interferometer}, so that numerical simulation of the transmission temporal shift of single Michelson interferometer is shown in Fig. \ref{Fig:result-many-two-interferometer}(a), and numerical simulation of the transmission temporal shift of two cascaded Michelson interferometers is shown in Fig. \ref{Fig:result-many-two-interferometer}(b). Then the sensitivity-enhanced factor $Enh$= $\frac{3 \; ms}{0.06 \; ms}=50$ due to the Vernier-effect, which is the same value as the value $Enh$= $\frac{2.00 \; ms}{2.00\; ms -1.96 \; ms}=50$ calculated from Eq. (\ref{enhanced_factor}). The results verify that the sensitivity for measuring velocity can be enhanced much more according to the calculated sensitivity-enhanced factor (\ref{enhanced_factor}) by choosing two interferometers with smaller differences in FRS. However, in practice, there is a compromise between the sensitivity and measurement accuracy. Because the smaller FRS difference will increase the FRS of the envelope, which introduces the difficulty in tracking the temporal shift of the envelope peak \cite{SHAO201573}. 

The numerical results show that the MWVA technique based on Vernier-effect is more sensitive than the TWVA technique despite how much time is used in the two techniques. In the scheme of the TWVA technique,  Eq.(\ref{inter_meter_final2}) indicates that time shift $\delta t =\frac{4\pi \tau ^{2} v}{\lambda \phi}$ can be effectively enhanced by increasing the measuring time. Meanwhile, the Fig. \ref{Fig:result-many-two-interferometer}. show that our scheme costs more time than the scheme of the single Michelson interferometer. Therefore, our work leads to a new issue: is the MWVA technique based on Vernier-effect more sensitive than the TWVA technique at the same measurement time? And further discussion is shown in the next section.
\begin{figure*}[t]
	\centering
\subfigure
{
	\vspace{-0.2cm}
	\begin{minipage}{9.2cm}
	\centering
	\centerline{\includegraphics[scale=0.699,angle=0]{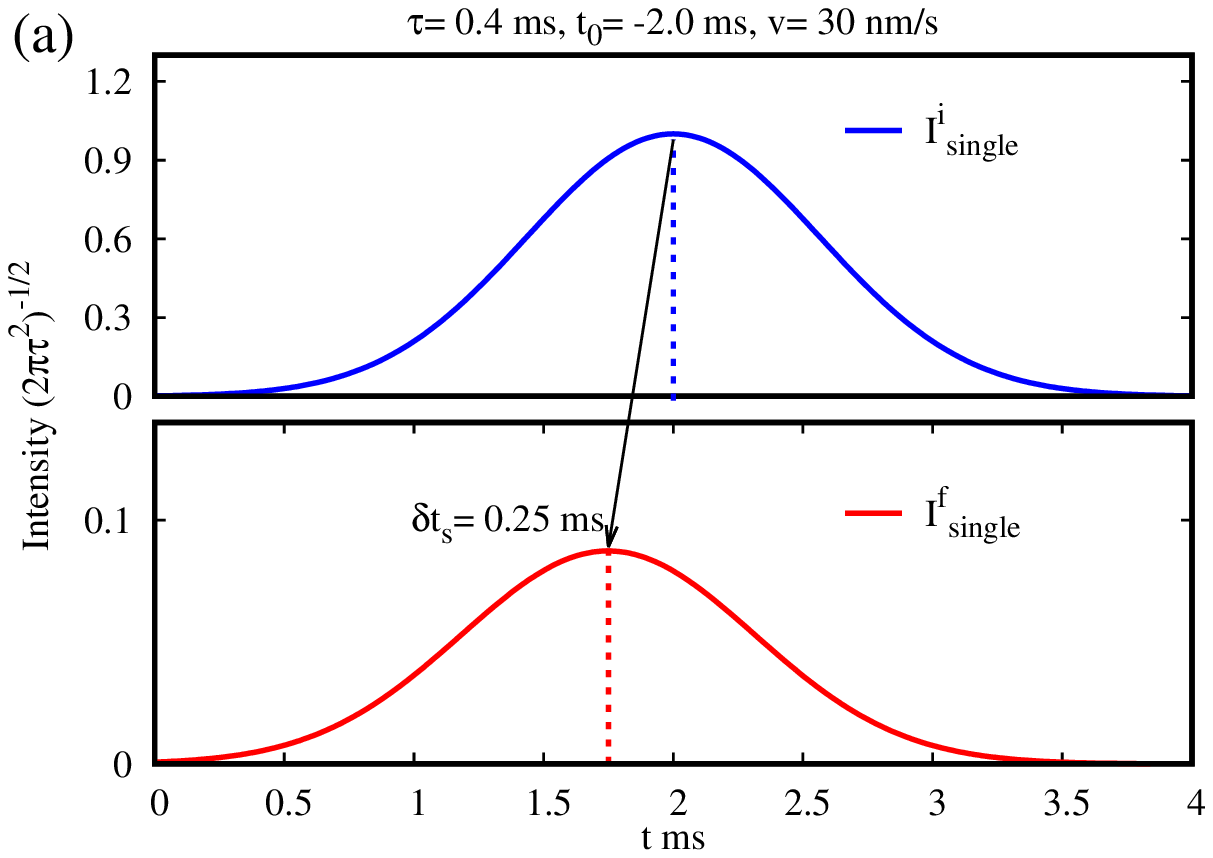}}
	\end{minipage}
}
\subfigure
{
	\vspace{-0.2cm}
	\begin{minipage}{8cm}
	\centering
	\centerline{\includegraphics[scale=0.699,angle=0]{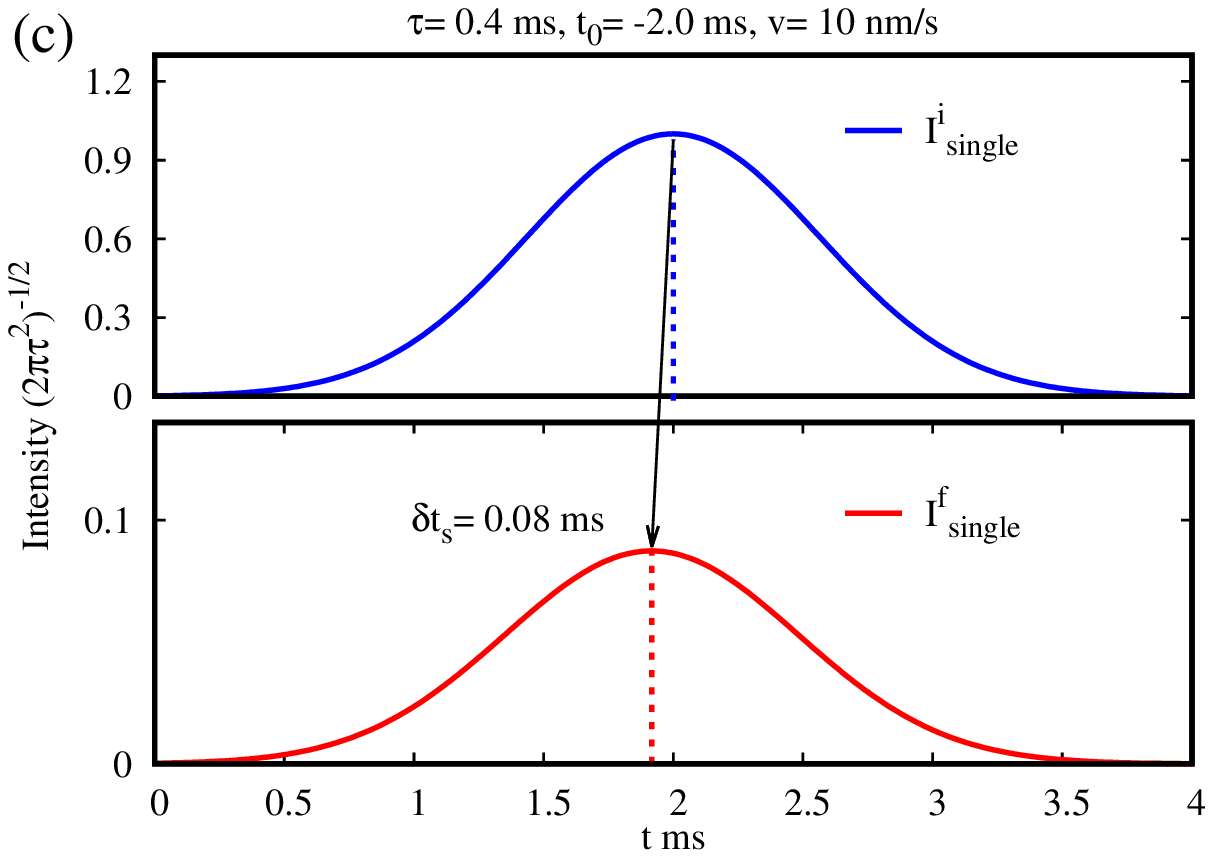}}
	\end{minipage}
}
\vspace{-0.2cm}

\subfigure
{
	\begin{minipage}{9.2cm}
	\centering
	\centerline{\includegraphics[scale=0.699,angle=0]{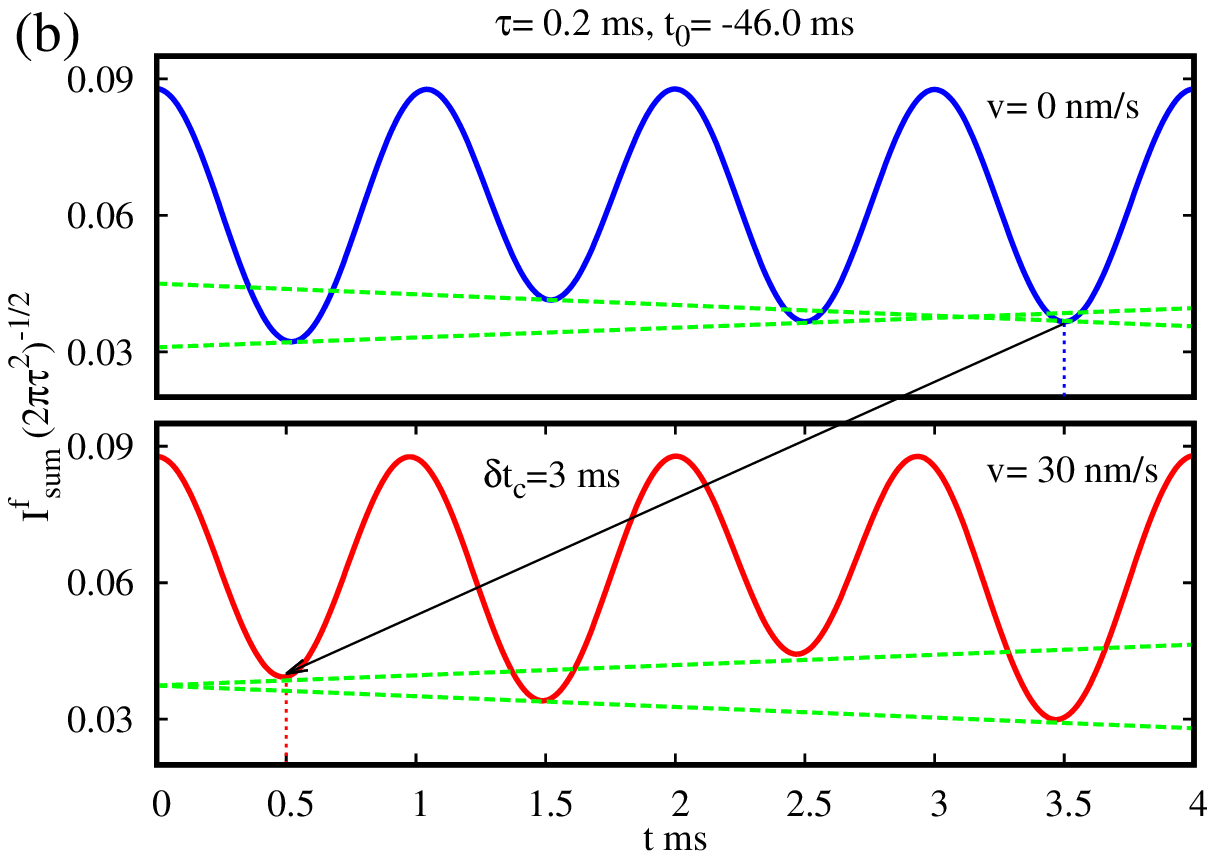}}
	\end{minipage}
}
\subfigure
{
	\begin{minipage}{8cm}
	\centering
	\centerline{\includegraphics[scale=0.699,angle=0]{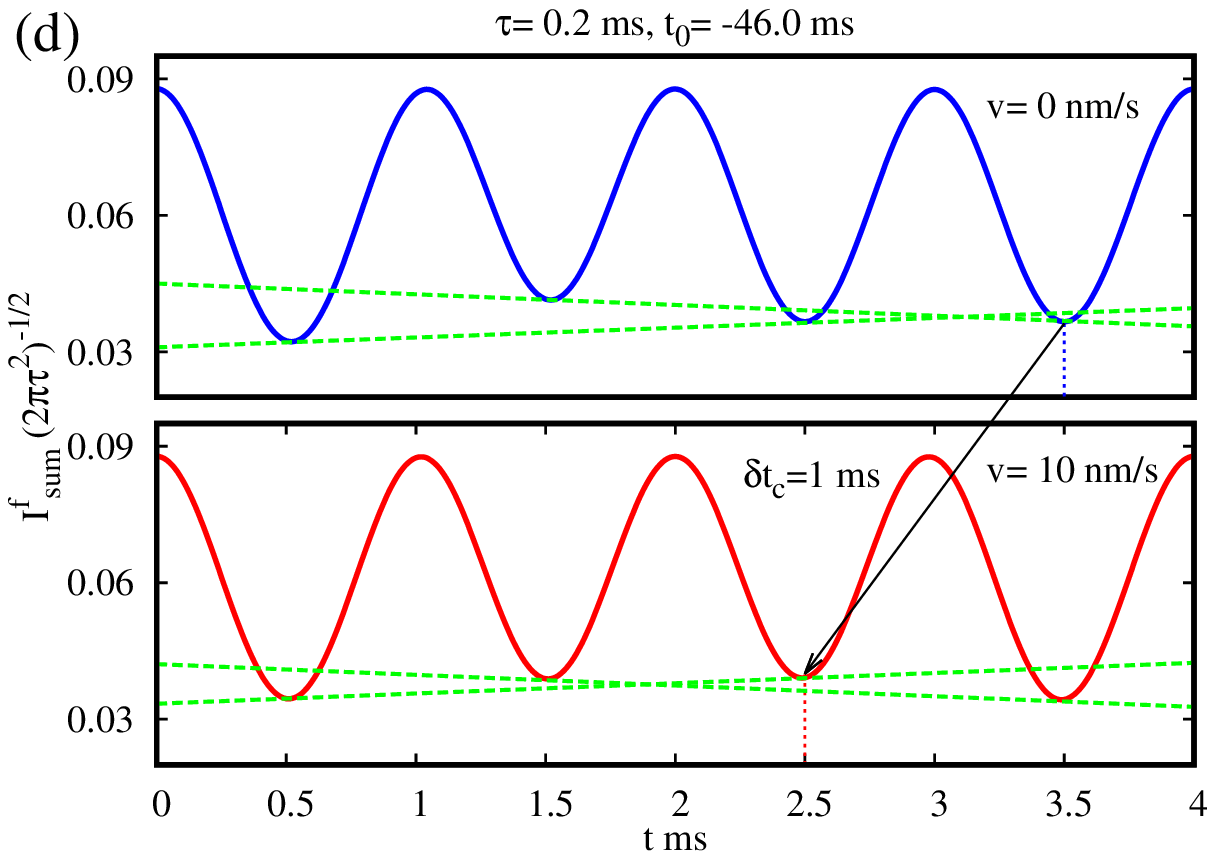}}
	\end{minipage}
}
\vspace{-0.2cm}
\vspace*{0mm} \caption{\label{Fig:result-many-two-interferometer-window30} 
Numerical simulation of the transmission temporal shift of (a) single Michelson interferometer and (b) two cascaded Michelson interferometers for measuring velocity $v$= 30 nm/s(left plane) and velocity $v$= 10 nm/s(right plane). The unit of the y-coordinate of each graph is $(2 \pi \tau ^{2})^{-1/2}$.}
\end{figure*}

\section{Further discussions and Numerical Results}
\label{SecIV}

\subsection{The sensitivity-enhanced factor at the same measurement time} 
\label{Secsensitivity-enhanced}

In this subsection, we will study the sensitivity-enhanced factor in the two schemes mentioned above at the same measurement time. Note that we can obtain the temporal shift in a certain and effective time window, and the rest time window is useless in the measurement. For example, When time is greater than 2 ms of the window in Fig. \ref{Fig:result-many-two-interferometer}(a), the measurement is invalid. 

In the next step, by choosing the appropriate meter in the TWVA technique and the MWVA technique, we compare the sensitivity of these techniques at the same measurement time. Note that we can obtain the temporal shift in a certain and effective time window. The simulation is shown in Fig. \ref{Fig:result-many-two-interferometer-window30}. The $I_{u}^{i}$ and $I_{d}^{i}$ are chosen with chosen post-selection phase $\phi=0.3  $ rad, the length $\tau$=0.2 ms of the individual Gaussian pulse, the parameter $t_{0}$=46.0 of the pulse $I_{u}^{i} $ (\ref{MVP_meter_initial_sliding}) and the pulse $I_{d}^{i} $ (\ref{MVP_meter_initial_fixed}), the free spectrum range $FSR_{fixed}$=2.0 ms of the lower interferometer and the free spectrum range $FSR_{sliding}$=1.96 ms of the upper interferometer in Fig. \ref{Fig:two-interferometer}. Then $I_{u}^{i}$ and $I_{d}^{i}$ are given by
$$ I_{d}^{i}=\left\{
\begin{array}{rcl}
\sum\limits_{m=0}^{N}
\frac{1}{\sqrt{2 \pi \times 0.2 ^{2}}}
e^{-(t+46-2.00m)^{2}/(2 \times 0.2 ^{2}}) &      & {t< 4ms }\\
0 &      & {t>4ms}
\end{array} \right. $$
$$ I_{d}^{i}=\left\{
\begin{array}{rcl}
\sum\limits_{m=0}^{N}
\frac{1}{\sqrt{2 \pi \times 0.2 ^{2}}}
e^{-(t+46-1.96m)^{2}/(2 \times 0.2 ^{2}}) &      & {t< 4ms }\\
0 &      & {t>4ms}
\end{array} \right. $$
In order to keep the same measurement time, the meter mode $I_{single}^{i}$ (\ref{inter_meter_initial}) of the schematic of the TWVA technique with the single Michelson in Fig. \ref{Fig:singel-interferometer} should be modulated with the length $\tau$=0.4 ms and the length $t_{0}$=-2.0 of the Gaussian pulse. The $I_{single}^{i}$ is given by:
$$ I_{d}^{i}=\left\{
\begin{array}{rcl}
\frac{1}{\sqrt{2 \pi \times 0.4 ^{2}}}
e^{-(t-2.0)^{2}/(2 \times 0.4 ^{2})} &      & {t< 4ms }\\
0 &      & {t>4ms}
\end{array} \right. $$

With the input of the $I_{single}^{i}$  of the single Michelson in Fig. \ref{Fig:singel-interferometer}, we can obtain the final meter $I_{single}^{f}$ (\ref{inter_meter_final2}) of measuring velocity $v$= 30 nm/s and $v$= 10 nm/s in Fig. \ref{Fig:result-many-two-interferometer-window30}(a) and Fig. \ref{Fig:result-many-two-interferometer-window30}(c). Under the same window of measurement time, the results of the detected intensity $I_{sum}^{f}$ of the two cascaded Michelson interferometers measuring velocity $v$= 30 nm/s and $v$= 10 nm/s based on Vernier-effect are shown in Fig. \ref{Fig:result-many-two-interferometer-window30}(b) and Fig. \ref{Fig:result-many-two-interferometer-window30}(d).

The method of how to chose the window of measurement time is explained as follows. The selected goal is to obtain the envelope through movement of $I_{sum}^{f}$ within the smallest possible measurement time window. In our work, we find that the intersection of straight lines determined by the interval troughs can track the shift of the envelope trough. More specifically, the shift of $I_{sum}^{f}$ can be calculated by tracking the shift of the trough, which is the nearest wave to the right of the intersection. Therefore, we call the above method the intersecting-point-tracing(IPT) method. By the IPT method, we obtain the shifts of the envelope $I_{sum}^{f}$ at measuring velocity $v$= 30 nm/s and $v$= 10 nm/s based on Vernier-effect are shown in Fig. \ref{Fig:result-many-two-interferometer-window30}(b) and Fig. \ref{Fig:result-many-two-interferometer-window30}(d). And the results show that the IPT method is effective and convenient for tracking the shifts.

We can obtain the effective sensitivity-enhanced factor $eEnh$ of measuring velocity $v$= 30 nm/s from Fig. \ref{Fig:result-many-two-interferometer-window30}(a) and Fig. \ref{Fig:result-many-two-interferometer-window30}(c). $eEnh$ is calculated by $eEnh=\frac{3.00 \; ms}{0.25 \; ms}=12$. In addition, the temporal shifts of measuring velocity $v$= 10 nm/s in Fig. \ref{Fig:result-many-two-interferometer-window30}(b) and  Fig. \ref{Fig:result-many-two-interferometer-window30}(d) conclude the same result $eEnh=\frac{1.00 \; ms}{0.08 \; ms}=12$. Though the effective sensitivity-enhanced factor $eEnh=12$ is smaller than the  sensitivity-enhanced factor $Enh=50$ calculated from  Eq. (\ref{enhanced_factor}), the sensitivity of the TWVA technique for velocity measurements is still enhanced by several orders of magnitude thanks to the Vernier-effect.

In the actual measurement process, for the temporal shift to be measurable, we require that $\delta t>\Delta t$, where $\Delta t$ is the resolution of the detector. Assuming that the shift $\delta t=  0.08 \; ms$ in Fig. \ref{Fig:result-many-two-interferometer-window30}(c) has reached the resolution $\Delta t$ and the $\delta t$ can not be measured. However the shift $\delta t=  1.0  ms>\Delta t$ in Fig. \ref{Fig:result-many-two-interferometer-window30}(d) can be effectively measured due to the sensitivity-enhanced of Vernier-effect. Therefore, the MWVA technique based on Vernier-effect can break the resolution limit of the detector of the TWVA technique for velocity measurements.

\subsection{The analysis of the signal-to noise ratio based on the Fisher information} 

\begin{figure}[t]
	\centering
\subfigure
{
	\vspace{-0.2cm}
	\begin{minipage}{8cm}
	\centering
	\centerline{\includegraphics[scale=0.699,angle=0]{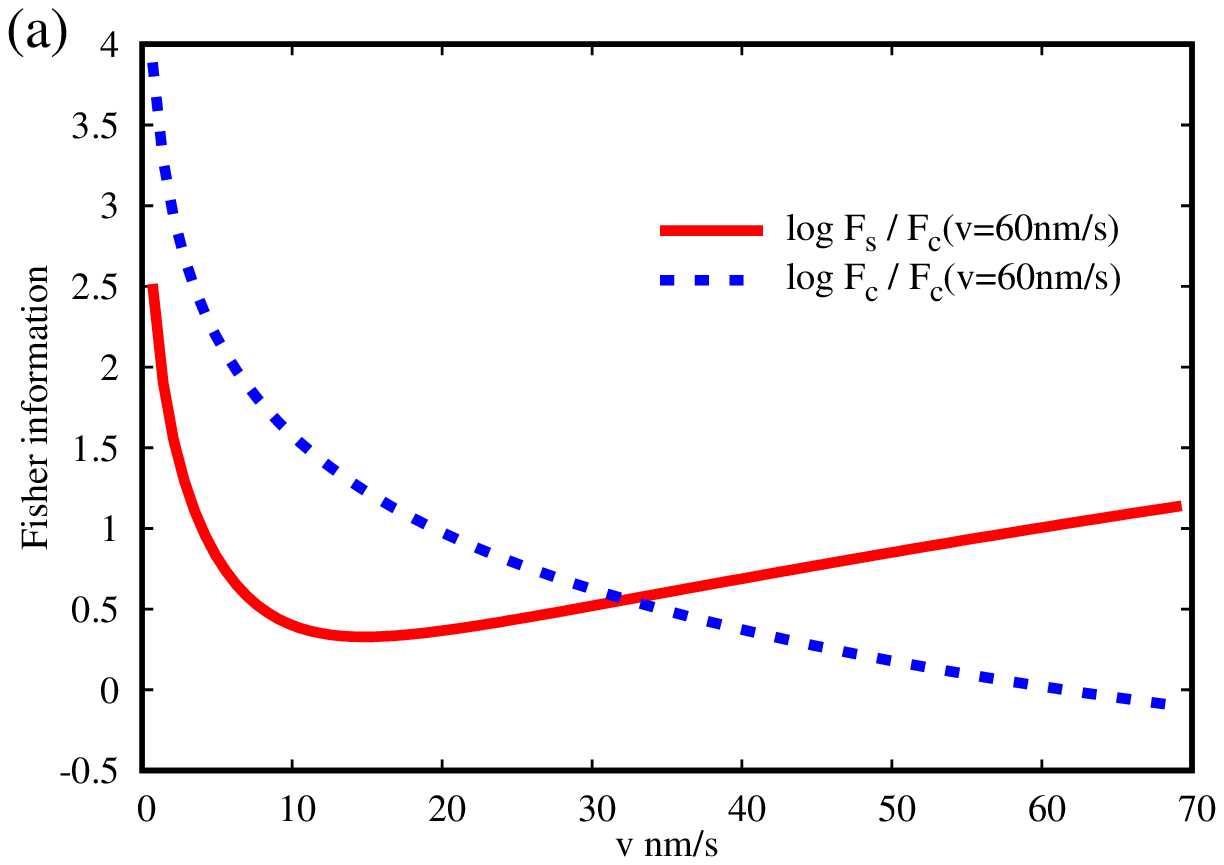}}
	\end{minipage}
}

\subfigure
{
	\begin{minipage}{8cm}
	\centering
	\centerline{\includegraphics[scale=0.699,angle=0]{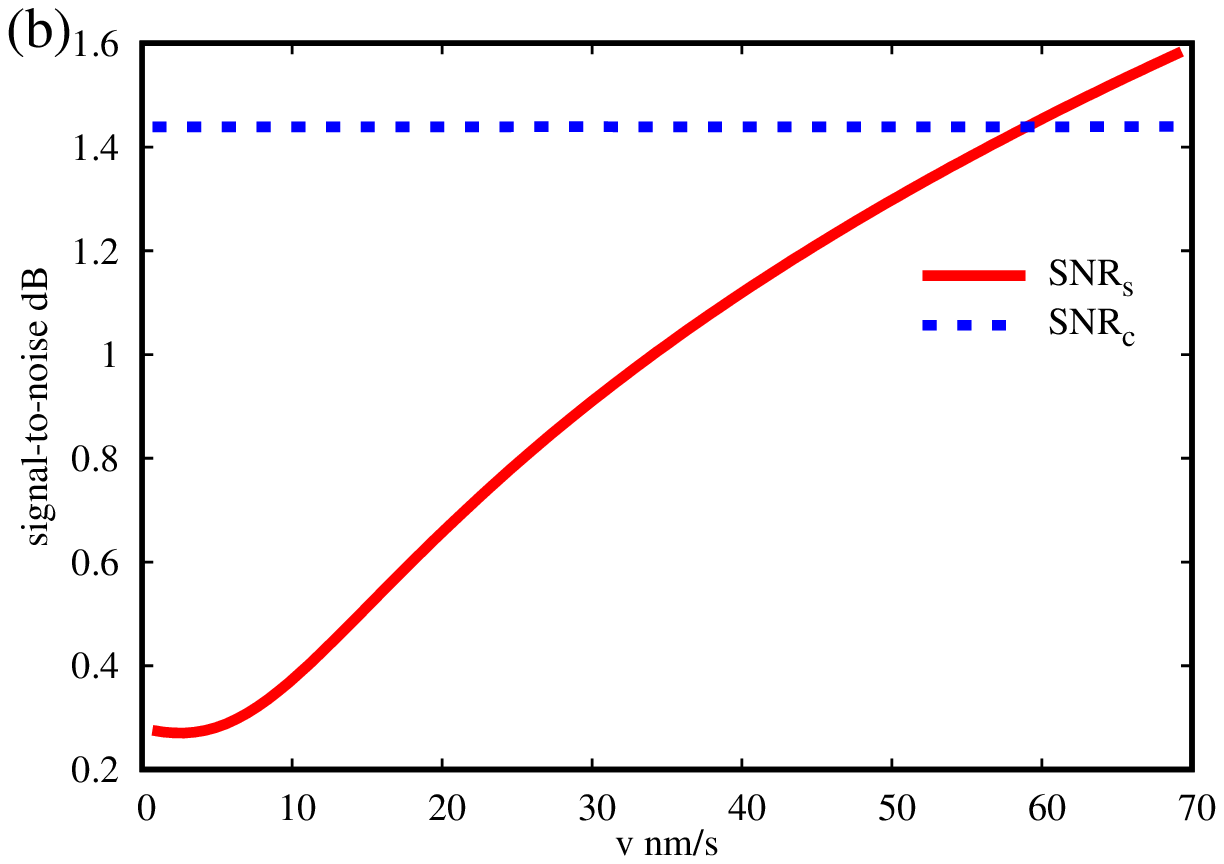}}
	\end{minipage}
}
\vspace*{0mm} \caption{\label{Fig:result-Fisher-SNR} 
The Fisher information (a) and the signal-to-noise (b) at different velocity with the traditional WVA technique and the modified WVA technique based on Vernier-effect.}
\end{figure}

In this subsection, by calculation the cram$ \rm \Acute{e}$r-Rao bound($CRB$)\cite{PhysRevE.88.042144,Pfister_2011}, we consider the fundamental limitation of velocity measurements with both the traditional WVA technique and the modified WVA technique based on the Vernier-effect. The $CRB$ is the fundamental limit in the minimum uncertainty for parameter estimation. And the $CRB$ is equal to the inverse of the Fisher information. In information theory, for a parameter $\Omega$ dependent probability distribution of a random variable $x$, $\textbf{P} (x|\Omega)$, the Fisher information\cite{2010Quantum} is defined as
\begin{eqnarray}
\label{Fisher_information_consept}
\textbf{F}(\Omega)
=N\int dx \; \textbf{P}(x|\Omega) [\frac{d}{d{\Omega}} \; {\rm ln}
\; \textbf{P}(x|\Omega)]^{2}
\end{eqnarray}
where $N$ represents that $N$ photons are sent through the interferometer. Then the $CRB$, which is the minimum variance $\Delta^{2}({\Omega})$ of an unbiased of ${\Omega}$ can be obtained as $CRB$=$1/\textbf{F}(\Omega)$. Next, we calculate $CRB$ and ${\rm SNR}$ in the TWVA technique and the MWVA technique based on Vernier-effect at the same measurement time $0<t<4 \;ms$(discussed in Sec. \ref{Secsensitivity-enhanced}). In our work, the parameter $\Omega$ is corresponding to the temporal shift $\delta t$.

In the scheme of Fig. \ref{Fig:singel-interferometer} with the traditional WVA technique, we chose the same pulse in Sec. \ref{Secsensitivity-enhanced} and obtain  the probability distribution of a random variable $t$:
$$ \textbf{P}_{s}=\left\{
\begin{array}{rcl}
\frac{(sin \phi)^{2}}{\sqrt{0.32\pi }}
e^{-(t-2.0+ \delta t)^{2}/0.32} &      & {t< 4ms }\\
0 &      & {t>4ms}
\end{array} \right. $$
and the Fisher information can be computed by
\begin{eqnarray}
\label{Fisher_information_singel}
\textbf{F}_{s}
=N\int dt \; \textbf{P}(t|\delta t) [\frac{d}{d{\delta t}}  \; {\rm ln}
\; \textbf{P}_{s}(t|\delta t)]^{2}
\end{eqnarray}
The sensitivity in determination of $\delta t=\frac{4\pi \tau ^{2} v}{\lambda \phi }$ is bounded by the square root of the minimum variance $\Delta(\delta t)=\sqrt{1/\textbf{F}_{s}}$. And the smallest resolvable velocity is determined by the signal-to-noise
\begin{eqnarray}
\label{Fisher_information_to_SNR}
SNR_{s}=\frac{\delta t_{s}}{\Delta (\delta t)}
=\frac{v}{\Delta v}=\frac{4 v \pi \tau^{2}\sqrt{\textbf{F}_{s}}}{\lambda \phi}
\end{eqnarray}
where $\delta t_{s}=\delta t$.

Meanwhile the Fisher information $\textbf{F}_{c}$ with the MWVA technique based on Vernier-effect in Fig. \ref{Fig:two-interferometer}. To keep the same  measurement times with the TWVA technique, we choose the input $I_{u}^{i}$ (\ref{MVP_meter_initial_sliding}) and $I_{d}^{i}$ (\ref{MVP_meter_final_fixed}) with the same parameters in Sec. \ref{Secsensitivity-enhanced}, then the Fisher information $\textbf{F}_{c}$ can be calculated by
\begin{eqnarray}
\label{Fisher_information_cascaded}
\textbf{F}_{c}
=N\int dt \; \textbf{P}(t|\delta t) [\frac{d}{d{\delta t}}  \; {\rm ln}
\; \textbf{P}_{c}(t|\delta t)]^{2}
\end{eqnarray}
where the probability distribution $\textbf{P}_{c}$ satisfies:
$$ \textbf{P}_{c}=\left\{
\begin{array}{rcl}
\sum\limits_{m=0}^{N}
\frac{1}{\sqrt{0.08\pi}}
[e^{-(t+46-2.00m+ \delta t)^{2}/0.08} &      &\\
 +e^{-(t+46-1.96m+ \delta t)^{2}/0.08} ]&      & {t< 4ms }\\
0 &      & {t>4ms}
\end{array} \right. $$
Therefore, the $SNR_{c}$ is given by:
\begin{eqnarray}
\label{Fisher_information_to_SNR_casaed}
SNR_{c}=\frac{  \delta t_{c}}{\Delta (\delta t)}
=\frac{12v}{\Delta v}=\frac{48 v \pi \tau^{2}\sqrt{\textbf{F}_{s}}}{\lambda \phi}
\end{eqnarray}
where $\delta t_{c}=12 \delta t$. Note that the $SNR_{c}$ with the TWVA technique will be enhanced with the factor $eEnh=12$ due to the Vernier-effect.

The complexity of the probability distribution $\textbf{P}_{s}$ and $\textbf{P}_{c}$ make it hard to analytically calculate the Fisher information. In our work, we numerically solve the integration (\ref{Fisher_information_singel}) and the integration (\ref{Fisher_information_cascaded}) for $N=\times 10^{6}$, and the results are shown in Fig. \ref{Fig:result-Fisher-SNR}(a). And on this basis, we calculate the  signal-to-noise at different velocities with the TWVA technique and the MWVA technique in Fig. \ref{Fig:result-Fisher-SNR}(b).

In our work, the cram$ \rm \Acute{e}$r-Rao bound($CRB$)\cite{PhysRevE.88.042144,Pfister_2011} states that the inverse of the Fisher information is a lower bound on the variance of any unbiased estimator of $\delta t$, by noting that $\Delta ^{2}(\delta t)$ characterizes the extent of the shot noise of the quantum measurement\cite{PhysRevA.102.042601}. 
Firstly, the intuitive explanation for the  Fisher Information is that the Fisher Information reflects the accuracy of our parameter estimation. The larger it is, the higher the accuracy of parameter estimation, that is, it represents more information. Fig. \ref{Fig:result-Fisher-SNR}(a) shows that the Fisher Information $\textbf{F}_{c}$ with our MWVA technique is larger than $\textbf{F}_{s}$ with the traditional TWVA technique, when our measured velocity is less than 32 nm/s (read from Fig. \ref{Fig:result-Fisher-SNR}(a)). However, with further increase of velocity $v$, our MWVA technique will become less efficient than the TWVA technique. As we have pointed out, our MWVA technique is an effective alternative to the TWVA technique, when the measurement is beyond the limits of resolution of the detector in Fig. \ref{Fig:singel-interferometer}.

Let us continue the discussion related to the signal-to-noise in  Fig. \ref{Fig:result-Fisher-SNR}(b). The results show that the signal-to-noise $SNR_{c}$ is larger than $SNR_{s}$ with the traditional WVA technique When our measured velocity is less than 58 nm/s (read from Fig. \ref{Fig:result-Fisher-SNR}(b)). Note that an interesting conclusion can be drawn from Fig. \ref{Fig:result-Fisher-SNR}. The Fisher information characterization is not fully equivalent to the ${\rm SNR}$ characterization with the increase of measurement velocity. And the similar conclusion is obtained by analyzing The WVA technique beyond the Aharonov-Albert-Vaidman limit in Ref \cite{PhysRevA.102.042601}. In addition, as the measurement velocity is smaller, the signal-to-noise ratio of our MWVA technique is improved more.

We notice that the time shift $\delta t_{c}$ in our  MWVA technique is less than 3.5 ms thanks to ensuring the same time measurement window. And the upper limit of the corresponding velocity measurement is 40 nm/s. Therefore, we can conclude that our MWVA technique is more efficient than the TWVA technique, because $SNR_{c}$ is larger than $SNR_{s}$ as $v<$ 40 nm/s in  Fig. \ref{Fig:result-Fisher-SNR}(b).

\section{Summary and Discussions}
\label{SecV}
In summary, using the feature of sensitivity-enhanced in Vernier-effect, we have proposed a modified-weak-value-amplification(MWVA) technique of measuring the mirror's velocity. Compared with the traditional-weak-value-amplification(TWVA) technique, we demonstrated sensitivity-enhanced and higher ${\rm SNR}$ by using two cascaded Michelson interferometers. In our work, the two cascaded interferometers are  composed of the similar optical structures. One interferometer with a fixed mirror acts as a fixed part of the Vernier-scale, while the other with a moving mirror acts as a sliding part of the Vernier-scale is for velocity sensing. By choosing the appropriate meter in our MWVA technique and in the TWVA technique to ensuring the same measurement time, we obtained numerical results of the sensitivity and the ${\rm SNR}$ in the two techniques. 

Our numerical results show that our MWVA technique is more efficient than the traditional one with 12 times sensitivity-enhanced and  higher ${\rm SNR}$. In addition, as the measurement velocity is smaller, the ${\rm SNR}$ of our MWVA technique is improved more. Note that our MWVA technique can propose a viable and effective alternative for the measurement out of the limit of the resolution in the TWVA technique. In addition,  by using the principles of the Vernier-effect, it is applicative and convenient to further improve the sensitivity and ${\rm SNR}$ in measuring other physical quantities with the WVA technique.

A few remarks are in order. We have proposed an intersecting-point-tracing(IPT) method  for tracking the shift of the meter with cascaded Michelson interferometers, the IPT method indicates that the intersection of straight lines which is determined by interval troughs can track the shift of the envelope trough. It is remarked that we have found that the fisher information characterization is not fully equivalent to the ${\rm SNR}$ characterization with the increase of measurement velocity. And the conclusion is the response to the attention in Ref \cite{PhysRevA.102.042601} from a purely theoretical interest.

In addition, our  MWVA technique can be realized beyond the Gaussian meter wave function. Note that the work\cite{Turek_2015} shows that the Hermite–Gaussian and Laguerre–Gaussian pointer states for a given coupling direction have advantages and disadvantages over the fundamental Gaussian mode in improving the ${\rm SNR}$. And the work\cite{PhysRevA.92.022109} indicates that the post-selected weak measurement scheme for non-classical pointer states(coherent, squeezed vacuum, and Schrodinger cat states) to be superior to semi-classical ones. Therefore, a MWVA technique with a non-Gaussian meter wave function  based on the Vernier-effect will be studied in our future works. Meanwhile, the relevant experiments are also being carried out gradually.

\begin{acknowledgments}
This study is financially supported by the National Key Research and Development Program of China (Grant No. 2018YFC1503705) and the Fundamental Research Funds for National Universities, China University of Geosciences(Wuhan) (Grant No. G1323519204). 

\end{acknowledgments}
\bibliography{reference}
\end{document}